# Modification of the law of mass action for homotypic interactions

Jérôme J. Lacroix

Graduate College of Biomedical Sciences, Western University of Health Sciences, 302 E 2nd st, Pomona, CA 91709, USA.

Correspondence to:

jlacroix@westernu.edu

**Abstract**

In its general definition, the law of mass action posits that, in systems where multiple elements move randomly, the rate by which they physically interact is proportional to the product of their densities. This law predicts the rate of elementary chemical reactions and many other phenomena in biology, physics, and social sciences. Using combinatorial mathematics, this study shows that this law inherently overestimates the rate of homotypic interactions, i.e. whereby identical elements interact with each other. This overestimation depends on how many of them interact simultaneously (homointeractivity), and on the difference between the density of interactants and homointeractivity. When this difference is large, as in most systems, the rate expression given by the law of mass action for any homotypic interaction is corrected by dividing it by the factorial of homointeractivity. This modification should yield better agreement between theoretical predictions and experimental determinations of interaction rates in a wide range of natural phenomena.

# Introduction

The law of mass action is an empirical law developed by Cato M. Guldberg and Peter Waage between the years 1864 and 1879 and independently by Jacobus H. Van't Hoff in 1877. At the time, this law explained the rate of simple chemical reactions such as ester formation and hydrolysis[1-3]. In chemistry, this law states that the rate of elementary chemical reactions, i.e. occurring in a single step, is proportional to the product of the concentration of each reactant. The law of mass action further proposes that, when chemical reactions reach their equilibrium, the product of the concentrations of reactants over products raised to their respective stochiometric coefficient-power is a constant:

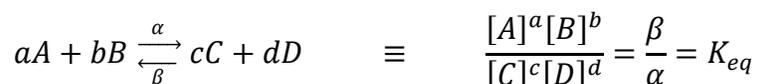

$$aA + bB \underset{\beta}{\overset{\alpha}{\rightleftharpoons}} cC + dD \quad \equiv \quad \frac{[A]^a[B]^b}{[C]^c[D]^d} = \frac{\beta}{\alpha} = K_{eq}$$

This fundamental law has laid down fundamental principles in chemistry, enzymology, and pharmacology. But because its basic principles extend to any system where elements randomly interact, this law also applies to many other fields including semi-conductor physics, epidemiology, and sociology[4,5].

A central tenet of the law of mass action is the intuitive assumption that, in a random interactome, the rate of interactions is a direct function of the frequency by which interactants meet or collide. The present study uses a simple thought experiment, followed by a rigorous mathematical approach, to demonstrate that the law of mass action systematically assumes a higher collision frequency when interactants are of the same nature. A general correction is proposed to account for this intrinsic bias.

**Results**

**A thought experiment reveals a bias in the law of mass action**

Let us consider a box (box 1) containing molecules $A$ and $B$, such that $[A] = [B] = c$; while a second box (box 2), of identical size and identical (non-zero) temperature, contains molecules $U$ at a concentration $[U] = 2c$, such that the total number of molecules in both boxes is the same. If the following elementary reactions occur:

$$A + B \xrightarrow{\alpha} AB \quad (box\ 1)$$

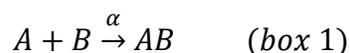

$$U + U \xrightarrow{\alpha} U_2 \quad (box\ 2)$$

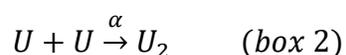

then, according to the law of mass action, the rate of $A - B$ interactions in box 1 is:

$$rate = \alpha[A][B] = \alpha cc = \alpha c^2$$

while in box 2, according to the same law, the rate of $U - U$ interactions is:

$$rate = \alpha[U]^2 = \alpha(2c)^2 = \alpha 4c^2$$

The law of mass action thus predicts that, if the rate constant $\alpha$ are equal in both cases, the interaction frequency - and thus the rate of product formation - should be four-fold greater in box 2 relative to box 1 (Figure 1). This seems counter intuitive. Indeed, both boxes have the same physical dimensions and contain the same number of molecules agitated at the same temperature. If the masses and sizes of molecules $A$, $B$, and $U$ are identical, the total number of intermolecular collisions counted over a sufficiently long duration should be the same in both boxes. In box 1, on average, only about half of those collisions would consist of potentially effective $A - B$ collisions and the other half would consist of ineffective $A - A$ or $B - B$ collisions. On the other hand, all

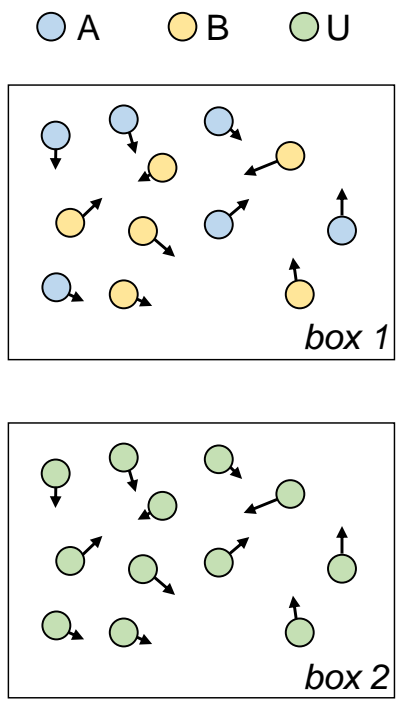

$$N_A = N_B = \frac{N_U}{2} = c$$

$$A + B \xrightarrow{\alpha} AB$$

$$rate = \alpha N_A N_B = \alpha c^2$$

$$U + U \xrightarrow{\alpha} U_2$$

$$rate = \alpha (N_U)^2 = \alpha (2c)^2 = \alpha 4c^2$$

**Figure 1: A thought experiment reveals a conundrum in the law of mass action.** Boxes 1 and 2 are identical and contain the same total number of molecules moved randomly by thermal agitation. In box 1, on average, half of all bimolecular collisions are between A and B, thus the rate of AB formation should be half that of UU formation in box 2. Yet, the law of mass action predicts a four-fold difference.

collisions in box 2 consist of potentially effective $U - U$ collisions. Thus, this thought experiment predicts twice as many potentially effective collisions in box 2, a 2-fold faster rate compare to box 1. Here, the term "potentially" refers to the fact only a fraction of intermolecular collisions has the necessary orientation and energy to form a molecular complex. However, the relative orientation of molecules at the time of collision is random and the collision energy depends on the relative velocities of molecules at the time of collision, which depends on the temperature-dependent Maxwell-Boltzmann distribution. For simplicity, only the frequency of collisions will be discussed, regardless of their ability to effectively form a stable complex.

**Homotypic vs. heterotypic bimolecular interactions**

Let us now use a combinatorial approach to determine the relative collisional frequencies between homotypic and heterotypic interactions. If box 2 has a volume $V$ and contains $N_U$ molecules, the relative frequency of bimolecular collisions between two $U$ molecules can be estimated from the probability $P_{UU}$ of exactly two molecules $U$ occupying the same small collisional volume, $v$ (this approach is an approximation only, as molecules are hereby considered "volumeless"):

$$P_{UU} = \binom{N_U}{2}\left(\frac{v}{V}\right)^2\left(1-\frac{v}{V}\right)^{(N_U-2)}$$

when $N_U \gg 2$, the binomial coefficient simplifies:

$$\binom{N_U}{2} = \frac{N_U!}{2!\,(N_U-2)!} = \frac{N_U(N_U-1)}{2!} \approx \frac{N_U^2}{2!}$$

the probability is thus approximated by:

$$P_{UU} \approx \frac{N_U^2}{2!}\left(\frac{v}{V}\right)^2\left(1-\frac{v}{V}\right)^{(N_U-2)} \qquad (1)$$

In box 1 (of same volume $V$), the probability $P_{AB}$ for having exactly one molecule $A$ and one molecule $B$ to occupy the same collisional volume $v$ is:

$$P_{AB} = \binom{N_A}{1}\left(\frac{v}{V}\right)\left(1-\frac{v}{V}\right)^{(N_A-1)}\binom{N_B}{1}\left(\frac{v}{V}\right)\left(1-\frac{v}{V}\right)^{(N_B-1)}$$

Which simplifies since $N_A = N_B = \frac{N_U}{2}$:

$$P_{AB} = \frac{N_U^2}{4}\left(\frac{v}{V}\right)^2\left(1-\frac{v}{V}\right)^{(N_U-2)} \qquad (2)$$

When comparing equations (1) and (2), we have the relation:

$$P_{AB} = \frac{1}{2} P_{UU}$$

This $\frac{1}{2}$ factor is also present when comparing the collision frequency between identical or different atoms or molecules in diluted gases or solutions[6]. Indeed, according to the collision theory, the frequency of bimolecular $A - B$ collisions in box 1, $Z_{AB}$ (s$^{-1}$ m$^{-3}$), is calculated by multiplying the relative mean velocity between colliding molecules, a collision cross-section surface, and the density of the molecules of interest in the system:

$$Z_{AB} = \frac{N_A N_B \pi (r_A + r_B)^2 (\langle v_A \rangle^2 + \langle v_B \rangle^2)^{\frac{1}{2}}}{V^2} \quad (3)$$

with $N_A$ and $N_B$, respectively the number of molecules $A$ and $B$, $r_A$ and $r_B$, respectively the radii of molecules $A$ and $B$ (m,) $\langle v_A \rangle$ and $\langle v_B \rangle$, respectively the mean speed of molecules $A$ and $B$ (m s$^{-1}$), and $V$, the volume of the system (m$^3$). On the other hand, the frequency of homotypic bimolecular $U - U$ collisions in box 2 is estimated using the following equation, obtained by replacing $\langle v_A \rangle$ and $\langle v_B \rangle$ by $\langle v_U \rangle$ and $r_A + r_B$ by, $d_U$, the diameter of molecules $U$, and halving the expression:

$$Z_{UU} = \frac{1}{2} \frac{N_U^2 \pi d_U^2 \sqrt{2} \langle v_U \rangle}{V^2} \quad (4)$$

Thus, when $\langle v_A \rangle = \langle v_B \rangle = \langle v_U \rangle$, $r_A = r_B = r_U$, and $N_A = N_B = \frac{N_U}{2}$, equations (3) and (4) lead to:

$$Z_{AB} = \frac{1}{2} Z_{UU}$$

which mirrors the results obtained using the combinatorial approach. Why does the expression used to calculate the frequency of homotypic bimolecular collisions includes a $\frac{1}{2}$ factor? One way to answer is to realize that the frequency of bimolecular interaction is a function of the number of order-independent combinations without repetitions of pairs of molecules that can form in the system: the higher the number of such pairs that can form, the higher the chance for them to form randomly. In box 1, the total number of such $A - B$ pairs is simply equal to the product $N_A N_B$. But in box 2, the total number of such $U - U$ pairs is equal to the binomial coefficient $\binom{N_U}{2} = \frac{N_U(N_U-1)}{2}$, which simplifies to $\frac{N_U^2}{2}$ when $N_U \gg 2$, hence the $\frac{1}{2}$ factor.

**Generalization for collisions between $h$ elements**

Using the same combinatorial approach, for heterotypic interactions, the probability that $h$ different interactants $A + B + C + \cdots$ collide simultaneously is:

$$P_{ABC\ldots} = \prod_{i=1}^{h} N_i \left(\frac{v}{V}\right) \left(1 - \frac{v}{V}\right)^{(N_i - 1)}$$

When the concentration of all interactant is the same ($N_i = N$), this simplifies to:

$$P_{ABC\ldots} = N^h \left(\frac{v}{V}\right)^h \left(1 - \frac{v}{V}\right)^{(hN - h)} \quad (5)$$

In contrast, in homotypic systems such as in box 2, the probability that $h$ identical interactants $U$ interact simultaneously is:

$$P_{UUU\ldots} = \binom{N_U}{h} \left(\frac{v}{V}\right)^h \left(1 - \frac{v}{V}\right)^{(N_U - h)}$$

When $N_U = N_A + N_B + N_C + \cdots = hN$, and $N_U \gg h$, this expression simplifies:

$$P_{UUU\ldots} = \frac{(hN)!}{h!\,(hN-h)!}\left(\frac{v}{V}\right)^h\left(1-\frac{v}{V}\right)^{(hN-h)} \approx \frac{h^h N^h}{h!}\left(\frac{v}{V}\right)^h\left(1-\frac{v}{V}\right)^{(hN-h)} \qquad (6)$$

Note that the ratio $\frac{P_{UUU\ldots}}{P_{ABC\ldots}}$ obtained by dividing equation (6) by equation (5) is equal to $\frac{h^h}{h!}$. In other words, when the assumptions $N_U = N_A + N_B + N_C + \cdots = hN$ and $N_U \gg h$ are valid (same total number of elements, the latter being much larger than the number of interacting elements), the probability of $h$ identical interactants to interact is $\frac{h^h}{h!}$ times greater than the probability of $h$ distinct interactants $A + B + C + \cdots$ to interact. By contrast, under the same assumptions, the law of mass action invariably leads to the conclusion that the rate at which $h$ identical interactants interact is $h^h$ times greater than the rate at which $h$ distinct interactants $A + B + C + \cdots$ interact. Thus, when the number of identical interactants is larger compared to $h$, the overestimation of collision frequency between interactants by the law of mass action is:

$$h^h \left(\frac{h^h}{h!}\right)^{-1} = h!$$

When the number of interactants $U$, $N_U$, is low compared to $h$, the approximation shown above in equation (2) is no longer valid. In this situation, the collision frequency of homotypic intermolecular reactions should be determined by calculating the binomial coefficients $\binom{N_U}{h}$. In the extreme case where $N_U = h$, the overestimation produced by the law of mass action is:

$$\frac{N_U{}^h}{\binom{N_U}{h}} = \frac{h^h}{\frac{h!}{h!\,(h-h)!}} = h^h$$

As expected, the limit of this error as $N_U$ increases toward positive infinity is equal $h!$ (Figure 2A), the value of the overestimation when $N_U \gg h$:

$$\lim_{n \to \infty} \frac{N_U{}^h}{\binom{N_U}{h}} = \lim_{n \to \infty} \frac{N_U{}^h h!}{\prod_0^{h-1}(N_U - h)} = h!$$

with $\lim_{n \to \infty} \frac{N_U{}^h}{\prod_0^{h-1}(N_U-h)} = 1$. The error made by approximated this overestimation by $h!$ is:

$$\left[\frac{N_U{}^h}{\binom{N_U}{h}} - h!\right]\frac{1}{h!} = \frac{N_U{}^h(N-h)! - N!}{h!\, N!}$$

The minimal value of $N_U$ necessary to keep this error below 10 % increases in a sub-logarithmic manner as $h$ increases (Figure 2B).

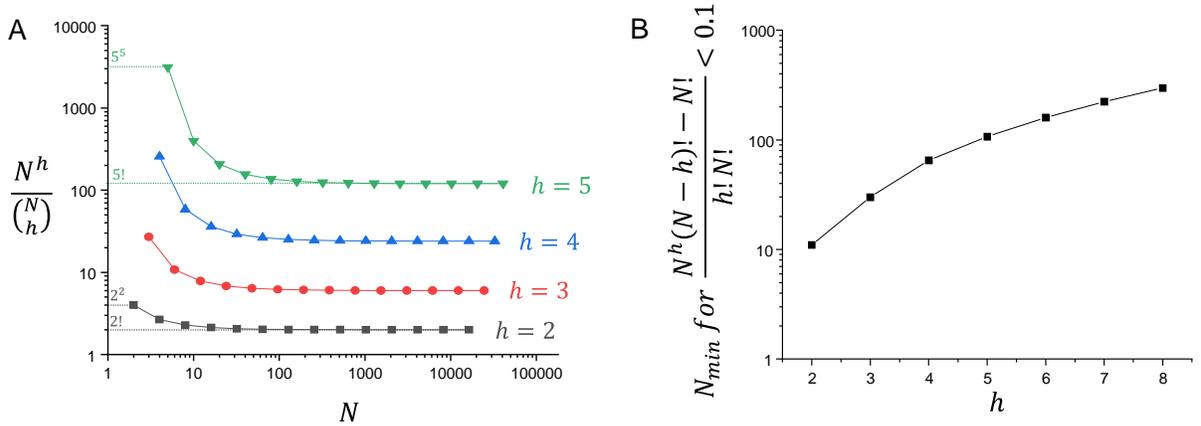

**Figure 2: Overestimation of homotypic interactions by the law of mass action.** (**A**) The overestimation, calculated by $\frac{N^h}{\binom{N}{h}}$ (Y-axis), is plotted as function of the number of homotypic interactants, $N$ (X-axis), and of the number of simultaneous interactions, $h$ (grey squares: $h = 2$, red circles: $h = 3$, blue up triangles: $h = 4$, and green down triangles: $h = 5$). (**B**) Plot showing the minimal value of $N$ ($N_{min}$) for which approximating the overestimation by $h!$ is less than 10% from its actual value.

**Discussion**

This study shows that the law of mass action systematically overestimates the rate at which identical elements interact. This overestimation depends on the number of interactants that are anticipated to simultaneously collide, a quantity defined here as "homointeractivity" and symbolized by $"h"$. When applied to chemistry, homointeractivity is equivalent to the molecularity of elementary chemical reactions. This overestimation varies from $h^h$ at low interactant densities and asymptotically reduces to $h!$ as the interactant density increases.

A general correction to the law of mass action for any interacting system can be easily implemented. The first step is to determine the presence of homotypic interactions and, if any, determine the value of its homointeractivity, $h$. When the number of interactants available in the system is larger compared to their homointeractivity, the correction consists of dividing the rate expression given by the law of mass action by the factorial of homointeractivity. For instance, for a hypothetical interaction with homointeractivity equal to 3, we have:

$$A + A + A \xrightarrow{\alpha} A_3$$

$$law\ of\ mass\ action\ rate\ =\ \alpha[A]^3$$

$$correct\ rate\ =\ \alpha[A]^3 \times \frac{1}{3!}$$

When the number of homotypic interactants is not very large compared to $h$, the best approach to obtain the reaction rate is to multiply the rate constant by the binomial coefficients according to the homointeractivity of the homotypic interaction:

$$A + A + A \xrightarrow{\alpha} A_3$$

$$law\ of\ mass\ action\ rate\ =\ \alpha[N_a]^3$$

$$correct\ rate\ =\ \alpha \binom{N_a}{3}$$

In chemistry, this conundrum has been either hidden in plain sight, perhaps because the overestimation of the rate of homotypic interaction does not modify the order of chemical reactions. This overestimation is a coefficient that is simply absorbed by the rate constants, which are, most of the time, empirically obtained. Nevertheless, a general law governing chemical equilibrium ought to place homotypic and heterotypic interactions on equal grounds. The law of mass action should thus be amended to account for its intrinsic bias.

Most chemical reactions happen in a succession of bimolecular elementary steps, hence the homointeractivity of each of these intermediate steps is not expected to exceed 2. A common reason for this paradigm is the exceedingly low frequency of collisions where more than 2 molecules collide, making "termolecular" chemical reactions very rare in nature. For instance, in a gas at normal temperature and pressure conditions, equation (6) shows that increasing $h$ by a single increment decreases collision probability by more than 3 orders of magnitude when $v$ is near the size of a small molecule. Since these bimolecular intermediate steps may or may not involve homotypic interactions, the homointeractivity of complex chemical reactions must therefore be empirically assessed based on the knowledge of the reaction mechanism.

The correction to the law of mass action proposed here should help obtain more consistent experimental rates and equilibrium constants between theory and experiments in a variety of interacting systems where the law of mass action applies.